# Improving Precipitation Estimation Using Multilinear Model Selection Algorithms

Ruhollah Nasiri[1*], Mohamad Sarajzadeh[2]

[1*] Department of Civil Engineering, Shiraz University, Shiraz, Iran. Email: R.nasiri@shirazu.ac.ir

[2] Department of Civil Engineering, Shahid Chamran University of Ahvaz, Ahvaz, Iran. Email: mserajzadeh@stu.scu.ac.ir

[*]Corresponding author

**Abstract**

High quality Quantitative Precipitation Estimation at high spatiotemporal resolution is crucial to many hydrologic/hydro-meteorological designs. Optimal Quantitative Precipitation Estimation of rainfall improves the accuracy of river and flash flood forecasts. In this study, we aim to merge multiple rainfall estimates including rain gauge, radar, Inverse Distance Weighting, Ordinary Co-Kriging, and Adaptive Conditional Bias Penalized Co-Kriging through two most common model selection techniques known as Least Absolute Shrinkage and Selection Operator and Bayesian Model Averaging. The methods were applied to the entire United States for a certain period. Statistical measures such as RMSE, ME, NSE, and Correlation Coefficient are used to investigate the accuracy and reliability of the estimation models. It is shown that both BMA and LASSO improve the precipitation estimation considering all ranges of rainfall observation included. However, OCK and CBPCK technique outperforms other methods in rainfall more than 10 mm. The IDW estimates show small bias, which results in a poor estimation, which is due to the limitation in using secondary variable radar. However, OCK and CBPCK address this problem by adding radar rainfall estimates as the second variable.

## 1. Introduction

The recent Industrializations and human Intrusion have made severe changes to the ecology and environment. Some studies have found several solutions to reduce the short- and long-term effects on the environment in terms of hydraulic engineering (Baharvand and Lashkar-Ara 2021; Rice et al. 2010; Kohankar Kouchesfehani et al. 2021). Different researchers have determined that hydraulic fracturing negatively affects human health and drives climate change. Some of the studies have considered the effects of industrialization and climate change in hydrology and water cycle such as prediction of drought using learning-based approaches (Hassanzadeh et al. 2020), sediment yield and scouring estimation (Lashkarara et al. 2021), flood hazard systems (Puttinaovarat and Horkaew 2020), and predicting the characteristics of hydraulic characteristics of different hydraulic phenomena (Baharvand et al. 2020).

Having a proper estimation for precipitation would have significant advantages for flood hazard systems. Various methods have been developed in measuring equipment (Samani et al. 2021) or numerical methods to investigate the flow discharge in a different flood or non-flood conditions. In addition to flow discharge estimation techniques, precipitation estimation is critical for accurate precipitation values over a case study. Maps of precipitation provide significant input for most hydrological and water resources planning models. However, precipitation estimations must be accurate to earn the full benefits from these models. Since precipitation is measured at a limited number of ground-based point rainfall data from sparsely positioned rain-gauge stations, producing a precipitation map from these point values requires an accurate estimation procedure. In order to estimate the precipitation from the spatial dataset, some methods such as Kriging (Phillips et al. 1992), Delaunay triangulations (Cohen and Randall 1998), inverse distance weighting (IDW, Bosch and Davis 1998), simple trend surface analysis (Gittins 1968), and Thiessen polygons have been used. These methods usually need a dense rain-gauge network with many stations (Adhikary et al. 2016). Economic, logistics, and geological factors apply significant restrictions on the rain-gauge network to be sparsely distributed in a field. As a result, point rainfall data are generally accessible from a limited number of gauge stations. These limitations increase the need to use suitable spatial estimation methods to obtain the spatial distribution of rainfall over a spatial dataset.

In recent years, weather radars as a secondary tool for precipitation estimation have become a popular tool. With the widespread use of weather radar systems, multisensory Quantitative Precipitation Estimation (QPE) using ground-based radar and rain gauge data is now a widely used method for precipitation estimation (Jozaghi 2021). Therefore, using weather radar, rain gauge data, and spatial estimation methods such as Kriging, Inverse Distance Weighting (IDW), Spline, Co-Kriging, different precipitation estimation models could be created over an area (Habib et al. 2012; Jozaghi et al. 2019, 2020, 2021; Kitzmiller et al. 2013).

Model selection is the most crucial part of any statistical analysis, and in fact, plays a central role in the pursuit of science in general (Hassanzadeh et al. 2020). Many authors have investigated this issue from both frequentist and Bayesian perspectives, and many techniques for selecting the "best model" have been proposed in the literature. The most widely used methods are Akaike Information Criterion (AIC, Akaike 1974), Exhaustive Search, F-tests for nested models, Stepwise, Backward, and Forward-selection, Mallows Cp, Bayes factors, Bayesian Model Averaging (BMA, Raftery 1993), Bayesian Additive Models for Location Scale and Shape (BAMLSS, Umlauf et al. 2018), Bayesian Information Criterion (BIC, Schwarz 1978). Data analysts always search to find a method that is coherent and general enough to tackle their problem. There are several sources of errors and uncertainties in hydrological modeling, including noisy measurements, model calibration, and simplifying assumptions (Clark et al. 2011; Elshall and Tsai 2014; Refsgaard et al. 2012; Renard et al. 2010). The stochastification of hydrological models will convert uncertainties into realistic uncertainty estimates for model predictions (Liu and Gupta 2007; Montanari and Koutsoyiannis 2012; Nearing et al. 2016). In the last two decades, multi-model methods based on Bayesian probability methods (Draper 1995) and information theory have become popular techniques to handle these uncertainties (Gupta et al. 2012).

Using Bayesian statistics in model selection have become a progressively essential method in the field of hydrogeologist such as groundwater modeling (Moazamnia et al. 2019; Rojas et al. 2010), reactive transport modeling (Lu et al. 2013), soil-plant-atmosphere modeling (Wöhling et al. 2015), and hydrological modeling (Marshall et al. 2005). Several investigations have been carried out to compare the performance of BMA relative to other methods, including log-linear models (Clyde and George 2000), linear regression (Baharvand et al. 2020; Fernández et al. 2001; Hoeting et al. 1999, 2002; Raftery et al. 1997), (Clyde and George 2000), graphical methods (Madigan et al. 1995; Madigan and Raftery 1994), logistic regression (Viallefont et al. 2001), semiparametric regression (Lamon and Clyde 2000), and binary regression (Fernandez et al. 2002). All these studies show that BMA has a better performance than alternative methods.

Another estimation method, which has been very popular since its introduction, proposed by Tibshirani (1996) is the Least Absolute Shrinkage and Selection Operator (LASSO). The LASSO minimizes the residual sum of squares subject to the sum of the coefficients' absolute value is less than a constant. Because of this constraint's nature, it tends to produce some precisely zero coefficients: therefore, this kind of model is an interpretable model.

This study aims to compare the performance of two standard model selection techniques BMA and LASSO. A multi-model merging module is developed, which merges the precipitation estimation techniques datasets. The model performances are compared to individual estimation sources using statistical measures using estimated and observed precipitation datasets. In addition, two widely used interpolation techniques, IDW, Ordinary Co-kriging, were used to generate the precipitation estimation spatially distributed maps. Furthermore, different merging approaches sensitivity analyses were done to present the best architecture for precipitation estimation based on BMA and LASSO models. The best model structure is developed and presented as a high-performance method to estimate the spatially distributed precipitation maps.

## 2. Materials and Methods

### 2.1. Data used

A fifteen day period starting from Oct 1, 2016 was used as the duration of the present study. Fig. 1 shows the radar-only precipitation for the study period. The reason for choosing this specific duration is that this period provides precipitation from storms in mainly three different regions including extreme amounts from Hurricane Matthew along the Atlantic Coast; significant amounts from a relatively weakly organized convective storm in the central US; and significant amounts from a coastal storm Pacific Northwest. The rain gauge data used are the hourly observations collected through the Hydrometeorological Automated Data System (HADS, Kim et al. 2009). The HADS is a real-time data acquisition, processing, and distribution system supporting the NWS's Flood and Flash Flood Warning programs.

Figure 2 shows the gauge station's spatially sparse locations, which cover 21000-gauge stations. The hourly radar QPE was used from the Multi-Radar Multi-Sensor Systems (MRMS; Zhang et al. 2011, 2016) with a resolution of 1 km. MRMS is a system of automated algorithms that integrate data from multiple radars, surface and upper-air observations, lightning detection systems, and satellite and numerical weather forecast models. The system generates a suite of 2D multisensory products for monitoring and short-term prediction of hail, wind, tornado, QPE, convection, icing, and turbulence. The radar precipitation estimates used in this work are the operational reflectivity-only MRMS QPE, referred to as Q3RAD (Cocks et al. 2017).

### 2.2. Precipitation map production

#### 2.2.1. Inverse Distance Weighting (IDW)

IDW was developed by the US National Weather Service in 1972 and is classified as a deterministic method. This method could be used for multivariate interpolation problems. Its general idea assumes that an unsampled point's attribute value is the weighted average of known values within the neighborhood (Lu et al. 2013). In this method, unknown spatial points value will be estimated using the neighbors' points with known values.

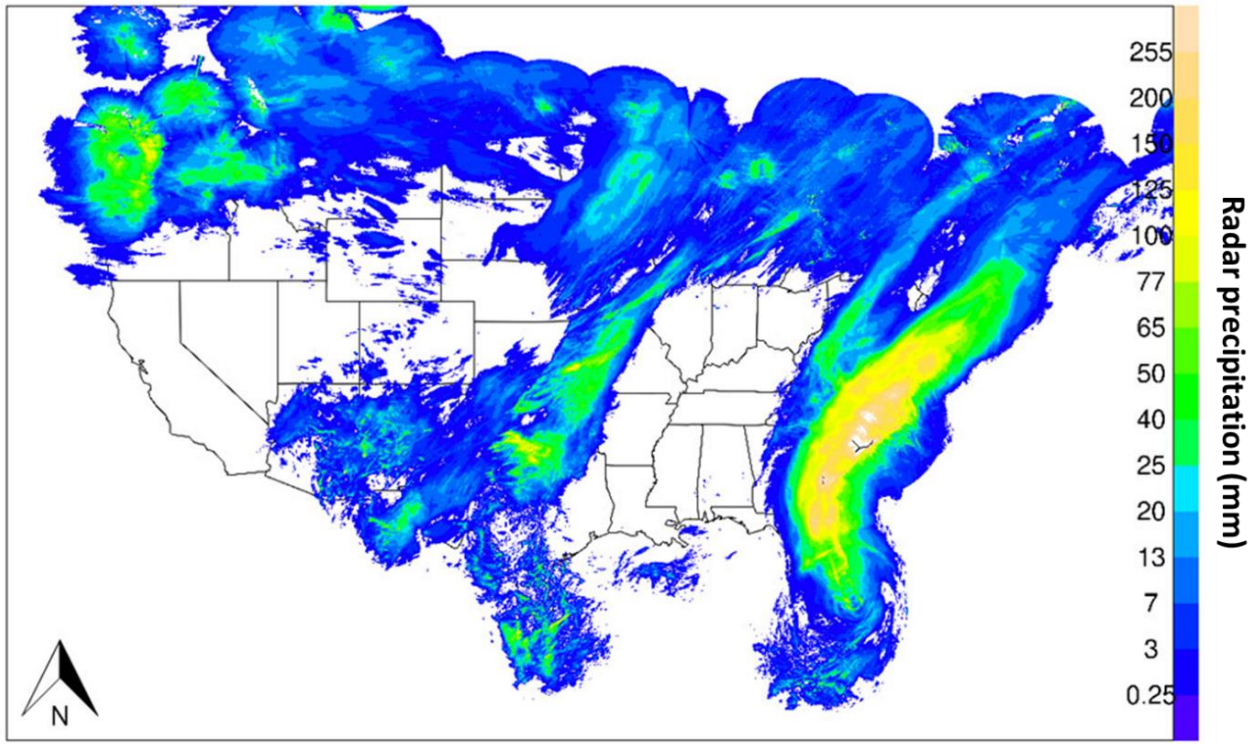

**Figure 1.** Radar rainfall map for Oct 1-15, 2016

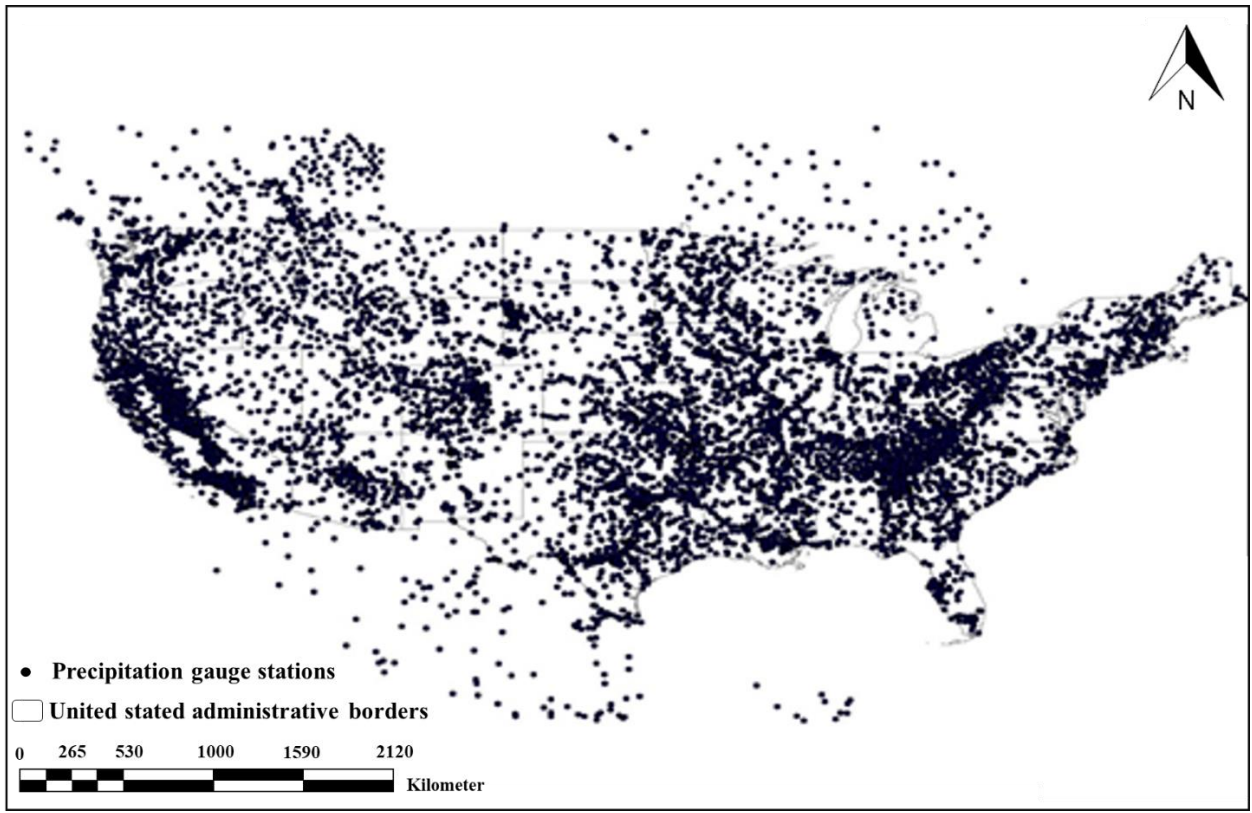

**Figure 2.** Rain gauge location map

The estimated value in each point of the study area will be the weighted sum of the values of $N$ gauge station points, in which $N$ is the total number of gauges with different distances from the target point. In terms of precipitation estimation problems, the IDW method estimates the unknown spatial rainfall data from the known sites adjacent to the anonymous site (Goovaerts 2000). The IDW formulas are given as Equations (1) and (2).

$$\hat{R}_P = \sum_{i=1}^{N} w_i R_i \tag{1}$$

$$w_i = \frac{d_i^{-\alpha}}{\sum_{i=1}^{N} d_i^{-\alpha}} \tag{2}$$

where $\hat{R}_P$ is the target (anonymous) rainfall data (mm); $R_i$ represents the rainfall data from observatory rainfall gauge stations (mm); $N$ refers to the number of rainfall gauge stations; $w_i$ is the weighting of each rainfall station; $d_i$ is the distance from each rainfall gauge station to the target point; $\alpha$ is the control coefficient. Several researchers studied the effect of control coefficient variations on the estimation accuracy of the IDW approach to estimate the precipitation (Simanton and Osborn 1980; Tung 1983). In the present study, $\alpha$ value is assumed 2.

### 2.2.2. Ordinary Co-kriging (OCK)

OCK method is a modification of the Ordinary Kriging (OK) method. The significant advantage of the OCK method is that it could be used for more than one variable rather than using only a single variable in the estimation process. The OCK method enhances the primary variable's estimation by using the secondary variable, assuming that the variables are correlated (Isaaks and Srivastava, 1989). In this study, rain gauge data and radar rainfall estimation are considered, respectively, as the primary and secondary variables in the OCK method. For the present study, the OCK estimator is derived in Equation (3) considering one secondary variable (radar precipitation data), which is cross correlated with the primary variable (rainfall gauge station data).

$$\hat{Z}_{OCK}(s_0) = \sum_{i1=1}^{n} w_{i1}^{OCK} Z(s_{i1}) + \sum_{i2=1}^{m} w_{i2}^{OCK} V(s_{i2}) \tag{3}$$

where $\sum_{i1=1}^{m} w_{i1}^{OCK} = 1$ and $\sum_{i2=1}^{m} w_{i2}^{OCK} = 0$, $\hat{Z}_{OCK}(s_0)$ is the estimated value of a primary variable at the target unsampled location $s_0$, $w_{i1}^{OCK}$ and $w_{i2}^{OCK}$ are the kriging weights associated with the sampling locations of the primary and secondary variables, Z and V, respectively. Also, n and m are the numbers of sampling points for the primary and secondary variables. The OCK weights are obtained by solving a system of (n+2) simultaneous linear equations that is derived in Equation (4).

$$\sum_{i1=1}^{n} \gamma_{zz}(s_{i1} - s_{j1}) w_{i1}^{OCK} + \sum_{i2=1}^{m} \gamma_{zv}(s_{i2} - s_{j1}) w_{i2}^{OCK} + \mu_1^{OCK} = \gamma_{zz}(s_{j1} - s_0) \tag{4}$$

$$\sum_{i1=1}^{n} \gamma_{vz}(s_{i1} - s_{j2}) w_{i1}^{OCK} + \sum_{i2=1}^{m} \gamma_{vv}(s_{i2} - s_{j2}) w_{i2}^{OCK} + \mu_2^{OCK} = \gamma_{vz}(s_{j2} - s_0) \tag{5}$$

$$\begin{cases} \sum_{i1=1}^{m} w_{i1}^{OCK} = 1 \\ \sum_{i2=1}^{m} w_{i2}^{OCK} = 0 \end{cases} \quad (6)$$

where $\gamma_{zv}(s_{i2} - s_{j1})$ and $\gamma_{vz}(s_{i1} - s_{j2})$ are the cross-variogram values between sampled $Z$ and $V$ values, $\mu_1^{OCK}$ and $\mu_2^{OCK}$ are the Lagrange multiplier parameters accounting for the two unbiased conditions. Figure 3 illustrates the map created using the OCK method for Hurricane Mathew from Oct 1, 2016, to Oct 15, 2016.

### 2.3. Model Selection Methods

Regression analysis is an approach to study the relationship between pairs of datasets. It is well documented that merging multiple estimation maps improves precipitation estimation with several sources of errors and uncertainties and creates a skillful assessment with better performance than every single estimation source. Two critical methodologies to account for uncertainty and combine these forecasts in statistics are the Bayesian and frequentist approaches. Frequentist techniques are characterized by studying parameter estimates' behavior in repeated sampling from the population. In contrast, Bayesian methods are described by updating the parameter's prior opinion based on the observed data (Contreras et al. 2018).

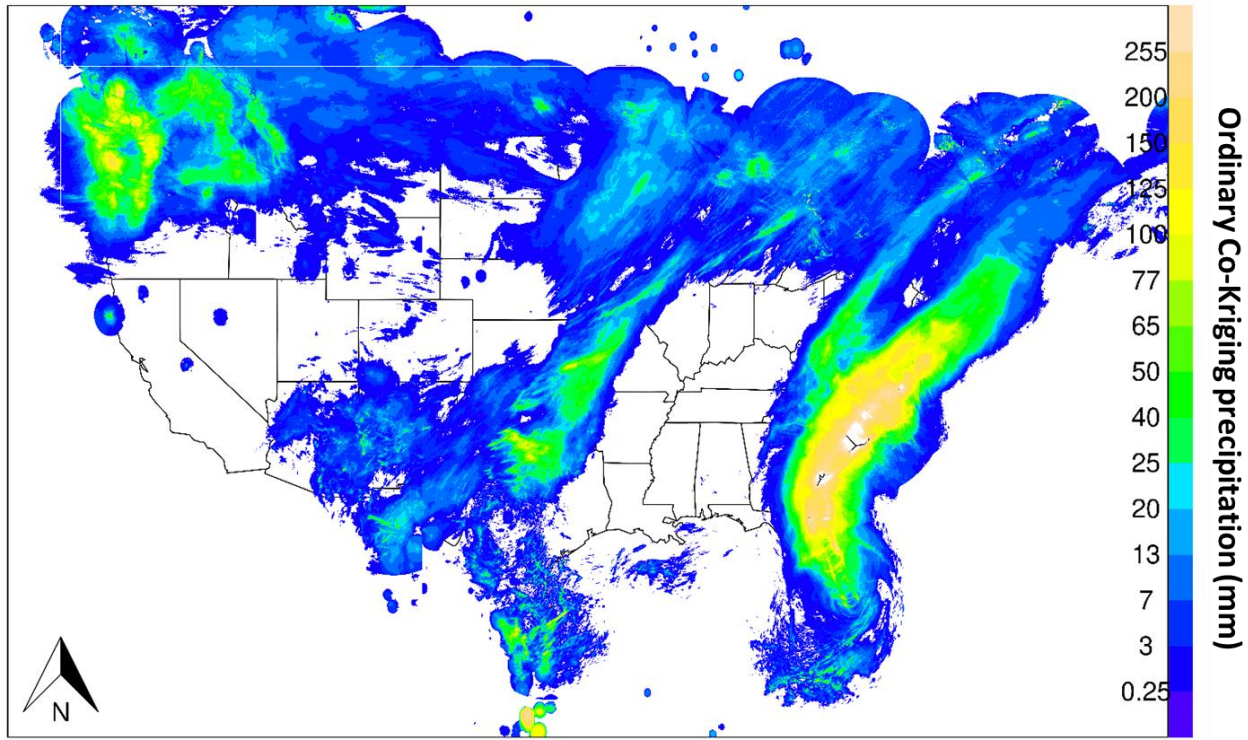

**Figure 3.** Ordinary Co-Kriging rainfall map for Oct 1-15, 2016

### 2.3.1. Bayesian Model Averaging

One of the most common techniques in model selection is Bayesian Model Averaging (BMA). Model averaging tries to combine the strengths of several models to improve inference and make a more skillful model (Kass and Raftery 1995). BMA provides a probability-based approach for considering multiple models in the presence of process and parameter uncertainty. Hoeting et al. (1999) provided a deep investigation in BMA with implementation details for selected model classes. A significant part of model averaging is that BMA was not designed as a technique for model selection, but rather it is used to combine posterior distributions.

Equation (7) shows a linear model equation to estimate the precipitation.

$$y = \alpha_i + X_i \beta_i + \varepsilon \quad \varepsilon \sim N(0, \sigma^2 I) \quad (7)$$

where y is observed streamflow; $\alpha_i$ is a constant; $X_i$ is a matrix of our estimation sources; $\beta_i$ represents the coefficients; $\varepsilon$ is a normal Independent and Identically Distributed (IID) error term with variance $\sigma^2$. However, if there are many predictors in a matrix X, it is crucial to find which predictors should be included in the model and how effective they are. BMA is a method developed by Bayesian statisticians to tackle this problem (George and McCulloch 1993). They found that averaging over multiple skillful models for all possible combinations of {X} and constructing a weighted average over all of them create a better predictive model than any single model under consideration. Thus, if we have several models {$M_1, M_2, \ldots, M_i, \ldots, M_n$}, it is not accurate to derive inference from a single model $M_i$. The posterior distribution of the parameter given the data is calculated using Equation (8) to measure the effect of a parameter in the presence of model uncertainty:

$$P(\beta|y, X) = \sum_{i=1}^{n} P(\beta|M_i, y, X) . P(M_i|X, y) \quad (8)$$

where n is the number of models which is equal to 2k, k is the number of estimation sources which is 4 in this study (Radar, OCK, CBPCK, and IDW), $P(\beta|y, X)$ is an average of the posterior distribution under each of the n models, with weight equal to the posterior model probability. This process is often complicated to tackle since it usually comprises many models, making it hard to compute the summation. Some integrals are not easy to calculate the target values (Kadane & Lazar 2004). Using the Markov chain Monte Carlo model composition (MC3) algorithm for Bayesian model averaging, we can overcome the former problem (Montanari and Koutsoyiannis 2012). The posterior model probability (PMP), which is a probability that model Mi is the true model, is given by Equation (9).

$$P(M_i|y, X) = \frac{P(y|M_i, X) . P(M_i)}{\sum_{j=1}^{n} P(y|M_j, X) . P(M_j)} \quad (9)$$

where $P(M_i|y, X) \quad i = 1, \ldots, n$ is the integrated likelihood of model, $P(M_i)$ is the prior probability that model $M_i$ is the true model. The integrated likelihood of the model is given by Equation (10).

$$P(y|M_i, X) = \int P(y|\beta_i, M_i, X) \cdot P(\beta_i|M_i) d\beta_i \quad (10)$$

where $\beta_i$ is the vector of parameters of the model $M_i$ (coefficients and variance), $\int P(y|\beta_i, M_i, X)$ is the likelihood, $P(\beta_i|M_i)$ is the prior density of $\beta_i$ under model $M_i$. According to Kass and Raftery (1995), predictors that have posterior effect probabilities between 50-75%, 75-95%, 95-99%, and 100% have weak, positive, strong, and robust evidence, respectively. When the posterior distribution is determined, the impacts of parameters on the observed rainfall will be calculated using posterior mean, posterior variance, and posterior effect probability, which is the summation of posterior model probability. Also, it measures the likelihood that a specific parameter is part of the true model, as shown in Equation (11) to (13).

$$E(\beta_1|y, X, \beta_1 \neq 0) = \sum_{A_1} \hat{\beta}_{1(i)}^2 \cdot P(M_i|y, X) \quad (11)$$

$$\begin{cases} Var(\beta_1|y, X, \beta_1 \neq 0) \\ \sum_{A_1} \left[ Var_{(i)} + \hat{\beta}_{1(i)}^2 \right] \cdot P(M_i|y, X) - E(\beta_1|y, X, \beta_1 \neq 0) \end{cases} \quad (12)$$

$$P(\beta_1 \neq 0|y, X) = \sum_{A_1} P(M_i|y, X) \quad (13)$$

where $\hat{\beta}_{1(i)}$ and $Var_{(i)}$ are maximum likelihood estimates and variance of $\beta_1$ under model i.

### 2.3.2. LASSO

Least Absolute Shrinkage and Selection Operator (LASSO) was first formulated by Robert Tibshirani (1996). It is a powerful method that performs two main tasks: regularization and feature selection. The LASSO method constrains the sum of the model parameters' absolute values; the sum must be less than a fixed value (upper bound). Thus, the method applies a shrinking (regularization) process to penalize the coefficients of the regression variables shrinking some of them to zero. During the features selection process, the variables that still have a non-zero coefficient after the shrinking process are selected to be part of the model. The goal of this process is to minimize the prediction error.

It is assumed that there is a dataset $(x^i, y_i)$, in which i = 1,2,…,N, and $x^i = (x_{i1}, x_{i2}, ..., x_{ip})^T$ are the predictor variables and $y_i$ is the responses. As in the usual regression set-up, we assume either that the observations are independent or that the $y_i$s are conditionally independent given the $x_{ij}$s. We assume that the $x_{ij}$ are standardized so that $\sum_i x_{ij}/N = 0$, $\sum_i x_{ij}^2/N = 1$.

Assuming $\hat{\beta} = (\hat{\beta}_1, \hat{\beta}_2, ..., \hat{\beta}_p)^T$, the lasso estimates $(\hat{\alpha}, \hat{\beta})$ using Equation (14).

$$(\hat{\alpha}, \hat{\beta}) = \arg\min \left\{ \sum_{i=1}^{N} (y_i - \alpha - \sum_j \beta_i x_{ij})^2 \right\} \quad (14)$$

Which is subjected to $\sum_j |\beta_i| \leq t$.

Here $t \geq 0$ is a tuning parameter. For the entire period (t), the solution for $\alpha$ is $\hat{\alpha} = \bar{y}$. We can assume without loss of generality that $\bar{y} = 0$ and hence omit $\alpha$.

The parameter $t \geq 0$ controls the amount of shrinkage that is applied to the estimates. Let $\hat{\beta}_j^0$ be the full least squares estimates and let $t_0 = \sum|\hat{\beta}_j^0|$. Values of $t < t_0$ will cause shrinkage of the solutions towards 0, and some coefficients may be exactly equal to 0. For example, if $t = t_0/2$, the effect will be roughly similar to finding the best subset of size $p/2$. Note also that the design matrix need not be of full rank.

## 3. Results and discussion

This section presents the merging results of interpolation techniques for Oct 1-15, 2016. This study used two different model selection techniques to find the best combination of these four estimation models (Radar, OCK, CBPCK, and IDW).

### 3.1. Bayesian Model Averaging of precipitation estimates

This section briefly presents the BMA model results for three days of accumulated precipitation. It investigates the model's accuracy based on the statistical measures between the observed gauge data and estimated values. Almost 21000 gauges all across the USA were used to create IDW, OCK, and CBPCK maps. Maps generated by each of mentioned methods with radar data were used as the predictors in the BMA model.

**Table 1.** Unconditional Error Statistic measures by BMA technique

|           | RADAR    | OCK      | CBPCK    | IDW      | BMA     |
|-----------|----------|----------|----------|----------|---------|
| RMSE (mm) | 2.7557   | 2.264    | 2.2517   | 5.2753   | 2.2205  |
| PRMSE %   | 19.42073 | 1.92062  | 1.38507  | 57.90757 | -       |
| ME (mm)   | 0.03546  | -0.02895 | -0.01099 | -0.82442 | 0.00076 |
| NSE       | 0.61497  | 0.74012  | 0.74293  | 0.411    | 0.75101 |
| CC        | 0.81742  | 0.8635   | 0.8662   | 0.12951  | 0.86623 |

Table (1) shows the results of BMA performance in an unconditional sense according to four different metrics Root Mean Square Error (RMSE), Mean Error (ME), Nash–Sutcliffe Efficiency (NSE, Nash and Sutcliffe 1970), and Correlation Coefficient (CC). The RMSE, MAE, NSE, and CC are calculated for the data set using Equations (15) to (18) (Jozaghi et al. 2018; Jozaghi and Shamsai, 2017).

$$RMSE = \sqrt{\frac{1}{n}\sum_{i=1}^{n}(Est_i - Obs_i)^2} \qquad (15)$$

$$ME = \frac{1}{n}\sum_{i=1}^{n}(Est_i - Obs_i) \qquad (16)$$

$$CC = \frac{\sum_{i=1}^{n}(Est_i - \overline{Est})(Obs_i - \overline{Obs})}{\sqrt{\sum_{i=1}^{n}(Est_i - \overline{Est})^2 \sum_{i=1}^{n}(Obs_i - \overline{Obs})^2}} \quad (17)$$

$$NSE = 1 - \frac{\sum_{i=1}^{n}(Obs_i - Est_i)^2}{\sum_{i=1}^{n}(Obs_i - \overline{Obs})^2} \quad (18)$$

where $Est_i$ is estimated value in point i, $Obs_i$ is observed value in point i, $\overline{Est}$ is the mean of estimated values in all n points, $\overline{Obs}$ is the mean of observed values in all points. According to Equation (16), a perfect match happens if NSE=1, If NSE=0, it shows that model estimations are as accurate as of the mean of the observed data. If NSE < 0, then the empirical mean is a better estimator than the model.

As shown in Table (1), the RMSE of BMA is 2.22, which shows the performance of BMA in an unconditional sense is better than all individual models.

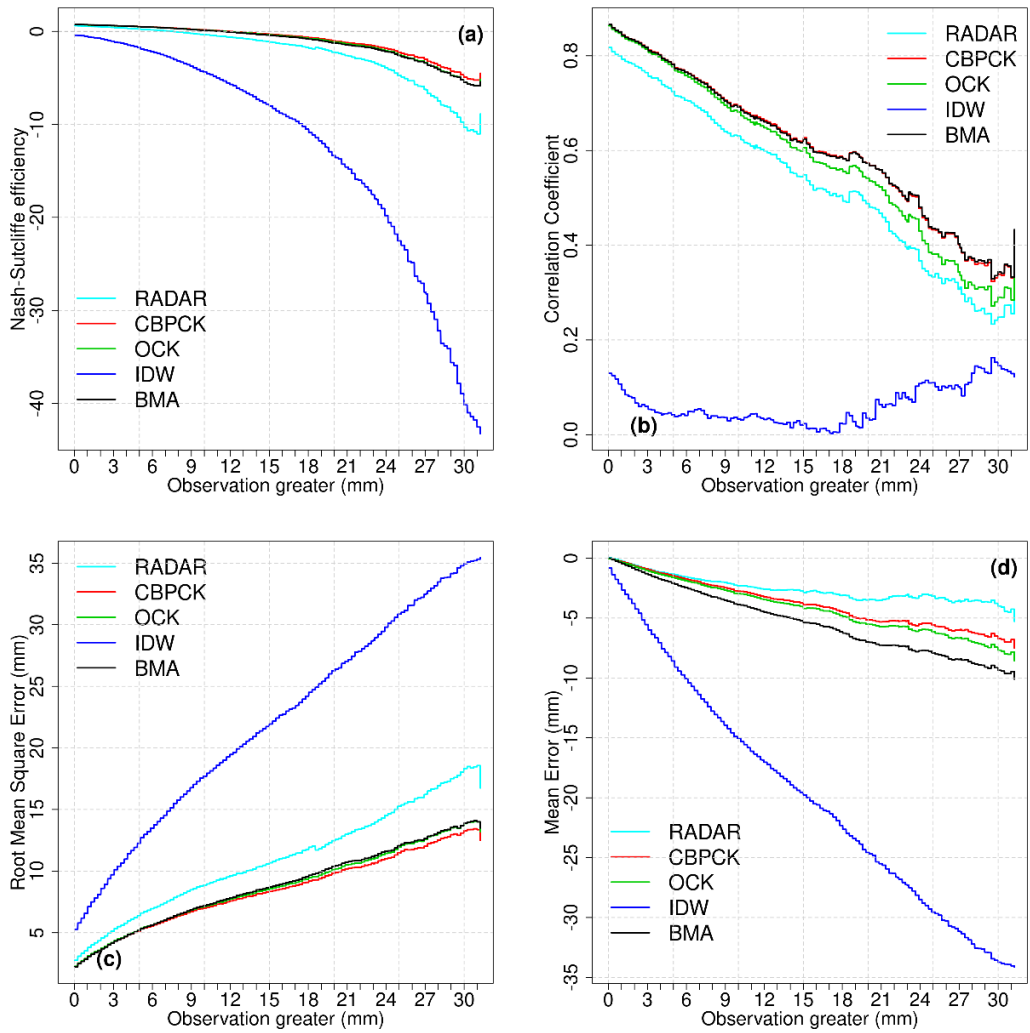

**Figure 4.** (a) Radar-only, adaptive CBPCK, OCK, IDW, and BMA estimations for Oct 1-15, 2016 versus a) NSE, b) CC, c) RMSE, d) ME

BMA decreases the RMSE of radar, OCK, CBPCK, and IDW almost 19.5%, 1.9%, 1.4%, and 57.9%, respectively. Also, we can see it also improves ME, NSE, and CC. To check the model's performance in a conditional sense, we plot NSE, CC, RMSE, and ME versus truth dataset. Figure (4) shows NSE, CC, RMSE, and ME of radar-only, OCK, CBPCK, IDW, and BMA estimates over the CONUS conditional on the verifying observed precipitation exceeding the amount shown on the x-axis. According to Fig (4) BMA has the best performance up to 15 mm precipitation, but in a larger amount of precipitation, CBPCK shows a better performance.

### 3.2. LASSO of precipitation estimates

Another model selection method used in this study is LASSO regression. We also check the results of this method in both conditional and unconditional sense. Table (2) presents the statistical measures of the LASSO method.

According to Table (2), the RMSE of LASSO is 2.223, which shows LASSO's performance in an unconditional condition is better than all individual models. LASSO decreases the RMSE of radar, OCK, CBPCK, and IDW almost 19.3%, 1.8%, 1.26%, and 57.86%, respectively. Besides, it can be seen that LASSO also improves ME, NSE.

Figure (5) shows NSE, CC, RMSE, and ME of radar-only, OCK, CBPCK, IDW, and LASSO estimations over the CONUS. According to Fig (5a), NSE of the LASSO is more extensive than all individual models up to 12 mm, but for rainfall values (rainfall >12mm), CBPCK improves the estimations. As we can see in the correlation coefficient plot, CBPCK and LASSO are more correlated with observations than the other models.

**Table. 2** Unconditional Error Statistic measures for LASSO approach

|  | RADAR | OCK | CBPCK | IDW | LASSO |
|---|---|---|---|---|---|
| RMSE (mm) | 2.7557 | 2.264 | 2.2517 | 5.2753 | 2.22322 |
| PRMSE % | 19.32341 | 1.80217 | 1.26598 | 57.85673 | - |
| ME (mm) | 0.03546 | -0.02895 | -0.01099 | -0.82442 | 0.00088 |
| NSE | 0.61497 | 0.74012 | 0.74293 | 0.411 | 0.7494 |
| CC | 0.81742 | 0.8635 | 0.8662 | 0.12951 | 0.86568 |

RMSE plot shows that LASSO has the best performance only in low precipitation up to 6 (mm) and after this amount of rainfall, OCK and CBPCK outperform the other methods. Due to Fig (5d), the radar data have a most negligible bias than all other models. Based on this figure, it could be argued that the LASSO approach has a negative bias with observations, which means LASSO underestimates our observations.

## 4. Conclusion

This paper used different interpolation techniques, created multiple precipitation maps, and then used them in our model selection techniques BMA and LASSO regression. According to tables. 1, 2 we can conclude that BMA slightly has a better performance than the LASSO regression model. Using BMA we can improve RMSE of radar, OCK, CBPCK, and IDW 19.5%, 1.9%, 1.4%, and 57.9%, respectively, whereas LASSO improves them 19.3%, 1.8%, 1.26%, and 57.86%, respectively. It can be seen the light differently in a conditional sense. Conditional BMA outperforms IDW and radar in all ranges of precipitation. It also improves OCK and CBPCK for rainfall smaller than 15 (mm). However, LASSO improves OCK and CBPCK up to 12 mm.

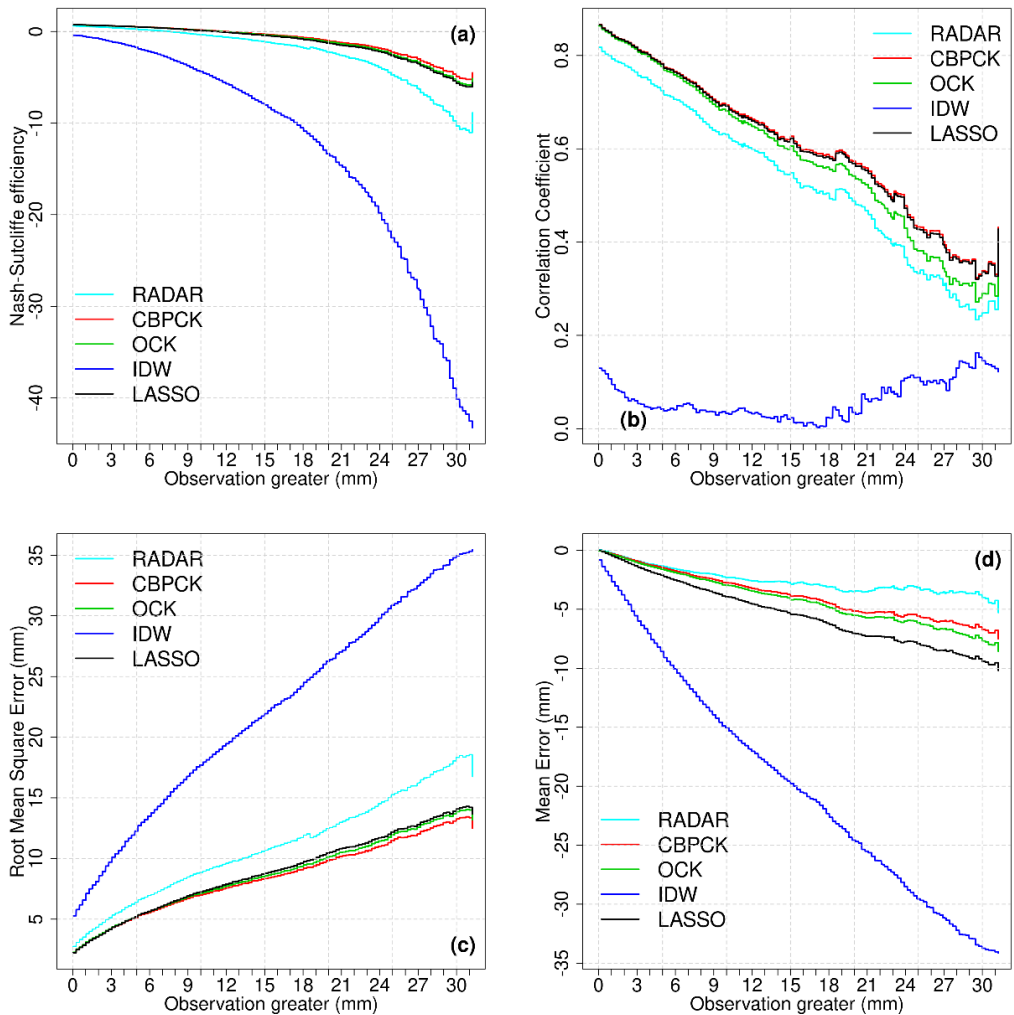

**Figure 5.** (a) Radar-only, adaptive CBPCK, OCK, IDW, and BMA estimations for Oct 1-15, 2016 versus a) NSE, b) CC, c) RMSE, d) ME

**Funding** This research received no external funding, and it was private research learning-based approach to assess and estimate the precipitation using novel techniques.

**Compliance with Ethical Standards**

**Conflicts of interest** The authors declare no conflict of interest.

*Advances in Water Resources*, 36, 36–50.

Renard, B., Kavetski, D., Kuczera, G., Thyer, M., and Franks, S. W. (2010). "Understanding predictive uncertainty in hydrologic modeling: The challenge of identifying input and structural errors." *Water Resources Research*, 46(5).

Rice, S. P., Lancaster, J., and Kemp, P. (2010). "Experimentation at the interface of fluvial geomorphology, stream ecology and hydraulic engineering and the development of an effective, interdisciplinary river science." *Earth Surface Processes and Landforms*, 35(1), 64–77.

Rojas, R., Kahunde, S., Peeters, L., Batelaan, O., Feyen, L., and Dassargues, A. (2010). "Application of a multimodel approach to account for conceptual model and scenario uncertainties in groundwater modelling." *Journal of Hydrology*, 394(3–4), 416–435.

Samani, Z. A., Baharvand, S., and Davis, S. (2021). "Calibration of Stage–Discharge Relationship for Rectangular Flume with Central Cylindrical Contraction." *Journal of Irrigation and Drainage Engineering*, 147(8), 06021006.

Schwarz, G. (1978). "Estimating the Dimension of a Model." *The Annals of Statistics*, 6(2).

Simanton, J. R., and Osborn, H. B. (1980). "Reciprocal-Distance Estimate of Point Rainfall." *Journal of the Hydraulics Division*, 106(7), 1242–1246.

Thiessen, A. H., (1911). "Precipitation averages for large areas." *Mon Weather Rev* 39(7):1082–1084

Tibshirani, R. (1996). "Regression Shrinkage and Selection Via the Lasso." *Journal of the Royal Statistical Society: Series B (Methodological)*, 58(1), 267–288.

Tung, Y. (1983). "Point Rainfall Estimation for a Mountainous Region." *Journal of Hydraulic Engineering*, 109(10), 1386–1393.

Umlauf, N., Klein, N., and Zeileis, A. (2018). "BAMLSS: Bayesian Additive Models for Location, Scale, and Shape (and Beyond)." *Journal of Computational and Graphical Statistics*, 27(3), 612–627.

Viallefont, V., Raftery, A. E., and Richardson, S. (2001). "Variable selection and Bayesian model averaging in case-control studies." *Statistics in Medicine*, 20(21), 3215–3230.

Wöhling, T., Schöniger, A., Gayler, S., and Nowak, W. (2015). "Bayesian model averaging to explore the worth of data for soil-plant model selection and prediction." *Water Resources Research*, 51(4), 2825–2846.

Zhang, J., Howard, K., Langston, C., Kaney, B., Qi, Y., Tang, L., Grams, H., Wang, Y., Cocks, S., Martinaitis, S., Arthur, A., Cooper, K., Brogden, J., and Kitzmiller, D. (2016). "Multi-Radar Multi-Sensor (MRMS) Quantitative Precipitation Estimation: Initial Operating Capabilities." *Bulletin of the American Meteorological Society*, 97(4), 621–638.

Zhang, J., Howard, K., Langston, C., Vasiloff, S., Kaney, B., Arthur, A., Van Cooten, S., Kelleher, K., Kitzmiller, D., Ding, F., Seo, D.-J., Wells, E., and Dempsey, C. (2011). "National Mosaic and Multi-Sensor QPE (NMQ) System: Description, Results, and Future Plans." *Bulletin of the American Meteorological Society*, 92(10), 1321–1338.17